\documentclass[twocolumn,aps,prl,superscriptaddress,showpacs]{revtex4}
\usepackage{hyperref}
\usepackage{graphicx}
\usepackage{amsmath}
\usepackage{epsfig}
\usepackage[utf8]{inputenc}

\newcommand{\ket}[1]{\mathop{\left| #1 \right\rangle}\nolimits}

\begin{document}
 \title{Probabilistic Cloning of Coherent States without a Phase Reference}

\author{Christian~R.~M\"{u}ller}
\affiliation{Max Planck Institute for the Science of Light, Erlangen, Germany}
\affiliation{Department of Physics, University of Erlangen-Nuremberg, Germany}

\author{Christoffer~Wittmann}
\affiliation{Max Planck Institute for the Science of Light, Erlangen, Germany}
\affiliation{Department of Physics, University of Erlangen-Nuremberg, Germany}

\author{Petr~Marek}
\affiliation{Department of Optics, Palack\'{y} University Olomouc, Czech Republic}%

\author{Radim~Filip}
\affiliation{Department of Optics, Palack\'{y} University Olomouc, Czech Republic}

\author{Christoph~Marquardt}
\affiliation{Max Planck Institute for the Science of Light, Erlangen, Germany}
\affiliation{Department of Physics, University of Erlangen-Nuremberg, Germany}

\author{Gerd~Leuchs}
\affiliation{Max Planck Institute for the Science of Light, Erlangen, Germany}
\affiliation{Department of Physics, University of Erlangen-Nuremberg, Germany}

\author{Ulrik~L.~Andersen}
\affiliation{Department of Physics, Technical University of Denmark, Kongens Lyngby, Denmark}
\affiliation{Max Planck Institute for the Science of Light, Erlangen, Germany}

\date{\today}

\begin{abstract}
We present a probabilistic cloning scheme operating independently of any phase reference. The scheme is based solely on a phase-randomized displacement and photon counting, omitting the need for non-classical resources and non-linear materials. In an experimental implementation, we employ the scheme to clone coherent states from a phase covariant alphabet and demonstrate that the cloner is capable of outperforming the hitherto best-performing deterministic scheme. An analysis of the covariances between the output states shows that uncorrelated clones can be approached asymptotically. An intriguing feature is that the trade-off between success rate and achieved fidelity can be optimized even after the cloning procedure.
\end{abstract}

\maketitle

In classical physics there are no fundamental limits to the performance of an amplifier or to the accuracy of copying a state of the system. 
The reason is that measurements can, in principle, gain the complete knowledge about the classical state, from which it is possible to generate arbitrarily many copies or a perfectly amplified version of the initial state. In general, this does not hold true in the quantum regime, since the laws of quantum mechanics forbid us to gain complete information about all aspects of reality on a single copy of a quantum state. This is widely known as the no-cloning theorem ~\cite{Wooters}. However, the no-cloning theorem does not prevent the creation of imperfect copies of a state~\cite{RevModPhys.77.1225}. These can, for instance, be obtained by amplification and subsequent splitting~\cite{OptQuCloning}. 

In general, the amplification of an unknown coherent state $\ket{\alpha}$ is accompanied by the addition of excess noise~\cite{louisell,haus,caves}. This noise is responsible for the fundamental bound for deterministic cloning, where the average fidelity of clones is limited to $F \leq \frac{2}{3}$. 
Nevertheless, the excess noise of amplifiers can be drastically reduced by relaxing the constraint of deterministic operation.
An ideal amplify-split cloner for coherent states is described by the two-step transformation $\ket{\alpha}\ket{0} \rightarrow \ket{\sqrt{2}\alpha}\ket{0} \rightarrow \ket{\alpha}\ket{\alpha}$. In the probabilistic regime, the amplification can for small amplitudes $|\alpha|$ be achieved with high accuracy as proposed in~\cite{T.C.Ralph2008, menzies_noiseless_2009, Marek2010, Fiurasek2009} and experimentally shown in~\cite{GrangierAmp,PrydeAmp,BelliniAmp,Usuga2010}. Yet, these approaches require perfect photon number detectors, single photon ancillary states and/or high-order non-linear interactions thus rendering the physical implementation challenging. Moreover, these approaches rely on additional key ingredients such as interferometric stability, perfect coincidences of single photon operations or a phase reference. The physical meaning and necessity of such a phase reference was discussed in an extensive debate~\cite{MoelmerPhase, MoelmerPhaseComment, MoelmerPhaseReply, WisemanPhase, FactistFictionistDialogue, RudolphSanders}.It has not been clarified what kind of quantum operations can be realized independently of any of these ingredients.

In this Letter, we demonstrate a phase covariant cloning scheme capable of probabilistically generating ideal clones of coherent states, without the above mentioned resources. We experimentally show that the cloner outperforms the bound set by the hitherto best-performing deterministic scheme~\cite{Sacchi2007} which is based on an optimal phase measurement. 
The cloner is of the amplify-split type and consists of solely elementary linear optical elements and a photon number resolving detector. The amplification is achieved probabilistically by first applying a phase randomized displacement to the input state and subsequently heralding the output depending on the result of a photon number threshold measurement on a part of the displaced state.

\begin{figure}
\includegraphics[width=8.5cm]{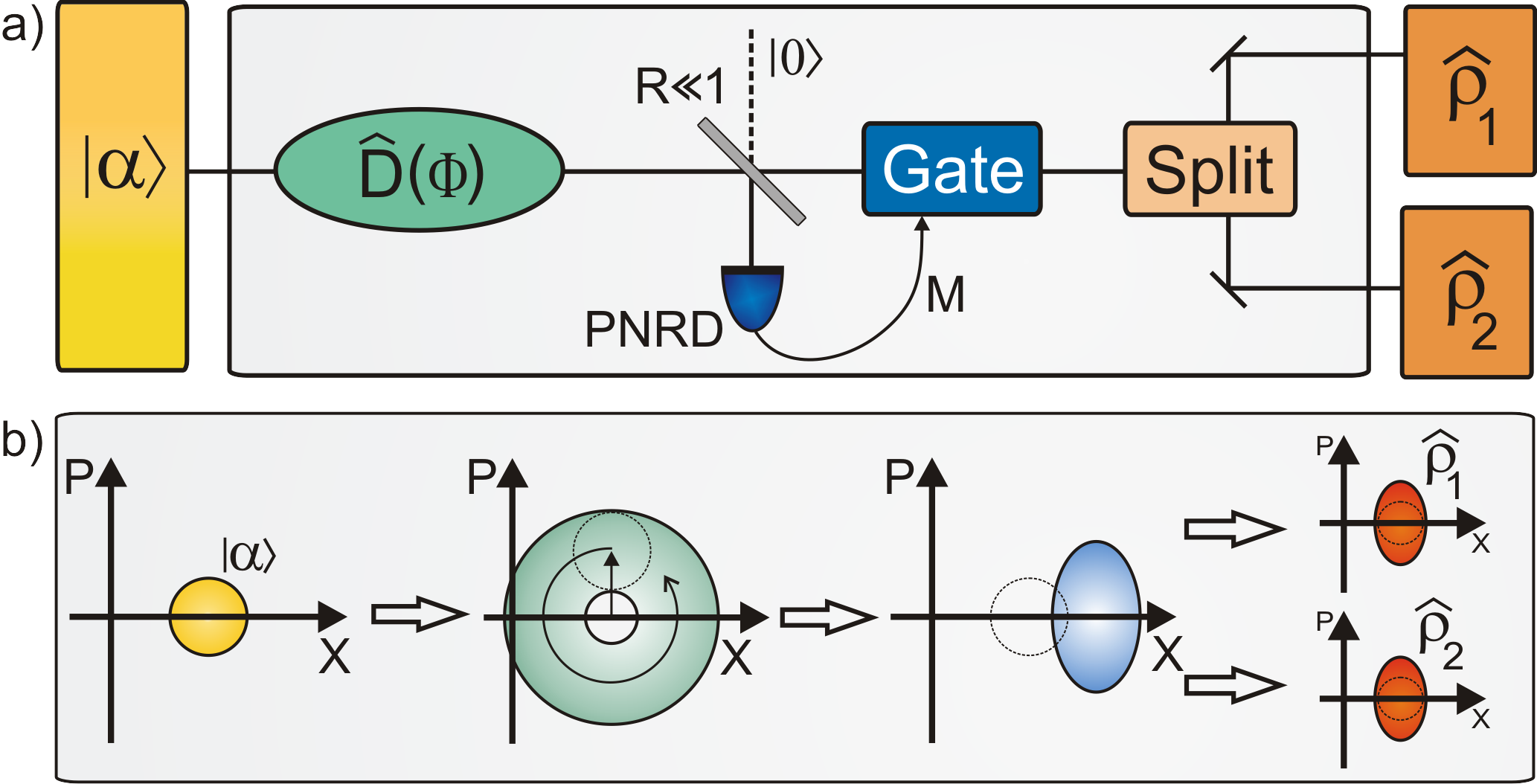}
\caption{(color online) a) Schematic of the cloning scheme. Two identical clones $\hat{\rho}_{1,2}$ are created by probabilistic amplification and subsequent symmetric splitting. b) Evolution of a coherent state in phase space during the cloning process.}
\label{Scheme}
\end{figure}

Our cloning strategy consists of three steps: displacement, heralding measurement and splitting, as sketched in Fig.~\ref{Scheme}(a).
First, the input state is randomly displaced in phase space ($\hat D(\Phi)$) according to a phase independent probability distribution $\Phi$. This is followed by probabilistic subtraction of photons via a photon number resolving detector (PNRD). Successful amplification is heralded whenever the detected number of photons surpasses a certain threshold value $M$. This strategy makes use of the classical correlations among the detected number of photons and the state's amplitude arising from the phase randomized displacement~\cite{Usuga2010}. The threshold parameter $M$ offers the possibility to tune the trade-off between output fidelity and success rate: Increasing $M$ will result in increased fidelity, however at the expense of lower success rate. In the final step the amplified state is split symmetrically to obtain the two copies of the input state. In contrast to other cloning protocols~\cite{Andersen2005, Sacchi2007}, our scheme disregards the phase information, as neither an external phase reference, which could be send along with the quantum states, nor an internal phase reference, e.g. a bright local oscillator, is provided. We merely need the definition of the input state's mode.

The phase insensitive annihilation $\hat{a}$ and creation $\hat{a}^{\dagger}$ operators constitute the fundamental building blocks of any quantum operation~\cite{Kim2008}. A specific class of operations that can be realized without a phase reference is described as $\hat{\rho} \rightarrow \sum_{n} \hat{A}_n \hat{\rho} \hat{A}_n^{\dagger}$, where the operators $\hat{A}_n$ are proportional to an arbitrary product of $\hat{a}$ and $\hat{a}^{\dagger}$. An elementary probabilistic amplification can be achieved without a phase reference for $\hat{A}_n = \hat{a}^m\hat{a}^{\dagger\,m}\,\delta_{mn}\,$\cite{Marek2010}, which amounts to adding and subtracting a specific number of photons $m$. However, a perfect coincidence between the additions and the subtractions is required for this operation. This constitutes a requirement in the particle picture, which is similar to providing a phase reference in the wave picture. In our scheme, this constraint is dropped  by replacing the single photon addition by the phase randomized displacement.

In the following, we discuss the phase covariant cloner for an alphabet with a fixed amplitude and continuous phase. In Ref.~\cite{Sacchi2007}, it has been shown that for such alphabets fidelities of at least $F \approx 0.85$ can be achieved deterministically, where the prerequisites are unit detection efficiencies, optimal phase measurements and feed-forward. An optimal scheme is not known, but this scheme is to the best of our knowledge the hitherto best-performing deterministic approach. We account for the amplification noise by choosing a displacement distribution with a fixed amplitude and random phase as sketched in Fig.~\ref{Scheme}(b). The phase randomized displacement ensures a phase insensitive operation and the fixed amplitude can be tailored to maximize the output fidelity. Insight into the cloning procedure can be gained by considering a weak coherent state $|\alpha\rangle \propto |0\rangle + \alpha|1\rangle$, with $|\alpha| \ll 1$. The state after the phase randomized displacement can be expressed as
\begin{equation}\label{}
    \hat{\rho} \propto \frac{1}{2\pi}\int_0^{2\pi} \hat{D}(|\alpha| e^{i\phi} \cdot x) |\alpha\rangle \langle \alpha|\hat{D}^{\dag}(|\alpha| e^{i\phi}\cdot x) d\phi,
\end{equation}
where $x$ is the ratio between the amplitude of the displacement and the original state's amplitude.
Subtracting a single photon transforms this state to
\begin{equation}\label{}
    \hat{\rho}_a \propto |\alpha|^2 \Big[ |0\rangle\langle 0| + x^2 (|0\rangle + 2\alpha|1\rangle)(\langle 0| + 2\alpha^* \langle 1|)   \Big], 
\end{equation}
which is a mixture of the vacuum state and the original state approximately amplified with an amplitude gain of two (i.e. four times the original power). From this state, four clones looking as
\begin{equation}\label{clonestate}
     \hat{\rho}_{clone} \propto |\alpha|^2 \Big[ |0\rangle \langle 0| + x^2 |\alpha\rangle\langle \alpha| \Big]
\end{equation}
can be generated and if the parameters are chosen such that $|\alpha|^2 \ll |x\alpha|^2 \ll 1$, perfect clones are obtained.
However, the usability of the cloner is by no means limited only to the extreme values required by the approximation; with the proper choice of $x$ and a multi-photon subtraction, we can achieve a respectable range of gains even for $|\alpha| \approx 1$.
Interestingly, for a fixed gain, the required value of $x$ is quasi constant for several numbers of subtracted photons, which allows for a delayed choice of $M$ and the trade-off between the success rate and the fidelity of the clones.

\begin{figure}
\includegraphics[width=8.5 cm]{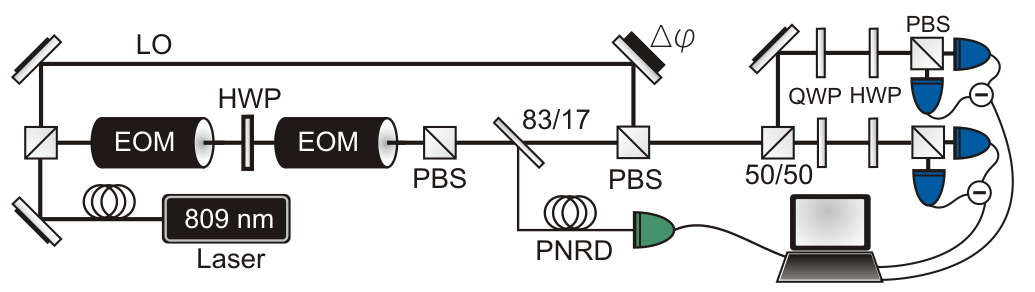}
\caption{(color online) Experimental realization of the cloning scheme.}
\label{setup}
\end{figure}

The experimental setup is sketched in Fig.~\ref{setup}. Our source is a grating-stabilized cw diode laser at $809\,\mathrm{nm}$ with a line width of $1\,\mathrm{MHz}$. After passing a fiber mode cleaner the beam is asymmetrically split into two parts, an auxiliary beam for the signal preparation and a local oscillator (LO) used only in the verification stage. The signal states are generated in time windows of $800\,\mathrm{ns}$ at a repetition rate of $100\,\mathrm{kHz}$. A combination of two electro-optical modulators (EOM) and a half wave plate (HWP) is used to generate and randomly displace a coherent state by transferring photons from the polarization mode of the auxiliary beam to the orthogonal signal polarization mode. A small portion ($\approx 17\,\%$) of the state is tapped off via an asymmetric beam splitter and sent to an avalanche photo diode (APD) operated in an actively gated mode. The dead time $(50\,\mathrm{ns})$ is much shorter than the pulse duration, allowing to employ the APD as a PNRD~\cite{CWittmann}. The heralded and effectively amplified state is finally split on a symmetric beam splitter to obtain the two clones. 
To quantify the fidelity between the input state and the clones, we perform full tomographies of both outputs. For this purpose, the amplified state is spatially mode matched with the LO on a polarizing beam splitter (PBS) yet before the state is split into the clones. Up to this stage signal and LO are still residing in orthogonal polarization modes.
The outputs of the beam splitter are directed to balanced homodyne detectors embedded in a Stokes measurement setup~\cite{PhysRevA.65.052306}. The combination of a HWP and a quarter wave plate (QWP) allows for the adjustment of the relative phase between the clone and the LO and therefore for simultaneous measurements of arbitrary quadratures at each output.
To enable tomography, the LO's phase is varied harmonically via a piezoelectric transducer to provide quadrature measurements of all phase angles. An accurate inference of the measured quadrature is provided by bright phase calibration pulses that are sent in between the signal states. The homodyne data and the number of detected photons are acquired simultaneously by a computer. Finally, we reconstruct the clones' density matrices employing a maximum likelihood algorithm~\cite{Hradil97, Lvovsky2004}. The homodyne detectors are only implemented to determine the performance of the scheme. In order to do so - unaffected by any imperfection of the verifying detection system - we assume unit quantum efficiency for the homodyne detectors and determine the performance not using the actual input amplitude but the following inferred value using $\eta_{HD_{1,2}}=1$: $|\alpha|^2 = |\alpha_{HD_1}|^2 + |\alpha_{HD_2}|^2 + \frac{1}{\eta_{PNRD}}|\alpha_{PNRD}|^2$, with $\eta_{PNRD}=63\pm3\,\%$.

The limited fidelity predicted by the no-cloning theorem is, in the case of an amplify-split cloner, due to the addition of excess noise in the amplification step. In our scheme, the excess noise is a remainder of the random displacement. This noise leads to classical correlations among the two clones which can be characterized by measuring the two-mode covariance matrix. Deterministic Gaussian strategies, for instance, add one unit of vacuum noise to the clones resulting in a uniform covariance of $0.5$. 
Probabilistic strategies are capable of generating ideal clones, having a vanishing covariance, asymptotically. However, in a realistic implementation a certain amount of excess noise is unavoidable.

\begin{figure}
\includegraphics[width=8.5cm]{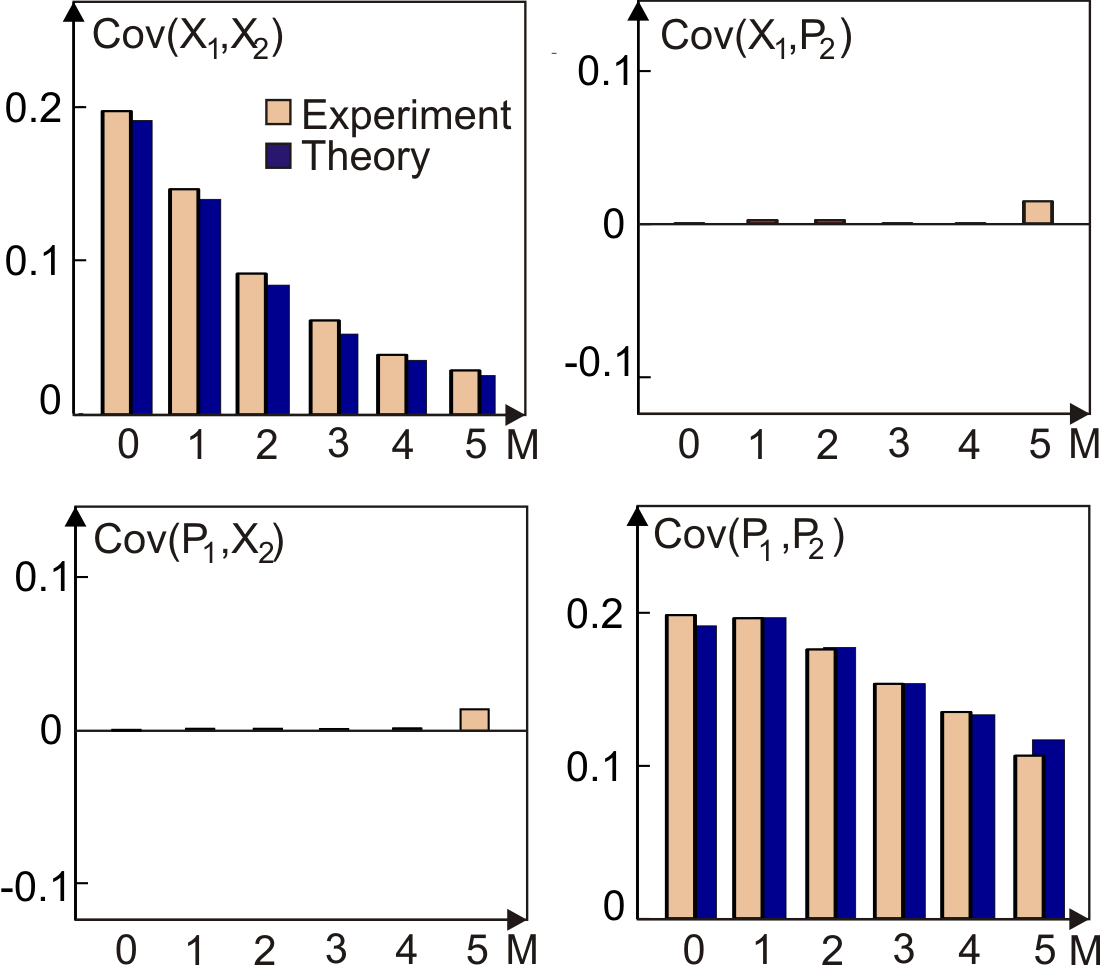}
\caption{(color online) Covariance matrix elements for the clones of a coherent state with initial amplitude $\alpha = 1.36$ for the randomly displaced state (M=0) and threshold parameters from one to five photon subtractions. The displacement parameter was chosen for all $M$ to be $x =0.5$. }
\label{CoVar_results}
\end{figure}

We measure the covariances for various threshold parameters and compare the results to the theoretical predictions. The results for a coherent state with amplitude $|\alpha| = 1.36$ excited along the X quadrature are presented in Fig.~\ref{CoVar_results}. In the experiment, we lock the LO's phase via a feedback loop and adjust the HWP and QWP at the detector stages to measure four different configurations to attain the in-phase terms $\mathrm{Cov}(X_1,X_2)$, $\mathrm{Cov}(P_1,P_2)$ as well as the out-of-phase terms $\mathrm{Cov}(X_1,P_2)$, $\mathrm{Cov}(P_1,X_2)$. The symmetric copies exit the cloner with identical phases, such that no out-of phase correlations are expected from the theory. This is confirmed by the experiment, apart from statistical fluctuations, which can become more pronounced with increasing threshold parameters and decreasing success rate.
Without heralding ($M=0$), the outputs have equal in-phase covariances. Heralding purifies the mixture by adding a bias to the high-amplitude parts of the mixture. Consequently, both in-phase covariances decrease with rising threshold. 
We find that the covariance along the direction of the state's excitation in phase space (for the state considered here: the $X$ quadrature) $\mathrm{Cov}(X_1,X_2)$ decreases faster than for the orthogonal quadrature $\mathrm{Cov}(P_1,P_2)$. In a simplified picture, the heralding process cuts off the low-amplitude part of the displaced state leaving only a segment of the initial ring-shaped displacement, which mainly spreads in the direction orthogonal to the excitation of the input state (see Fig.~\ref{Scheme}b)). We also find this behavior in the full theoretical model, which is in good agreement with the results.

\begin{figure}
\includegraphics[width=8.5cm]{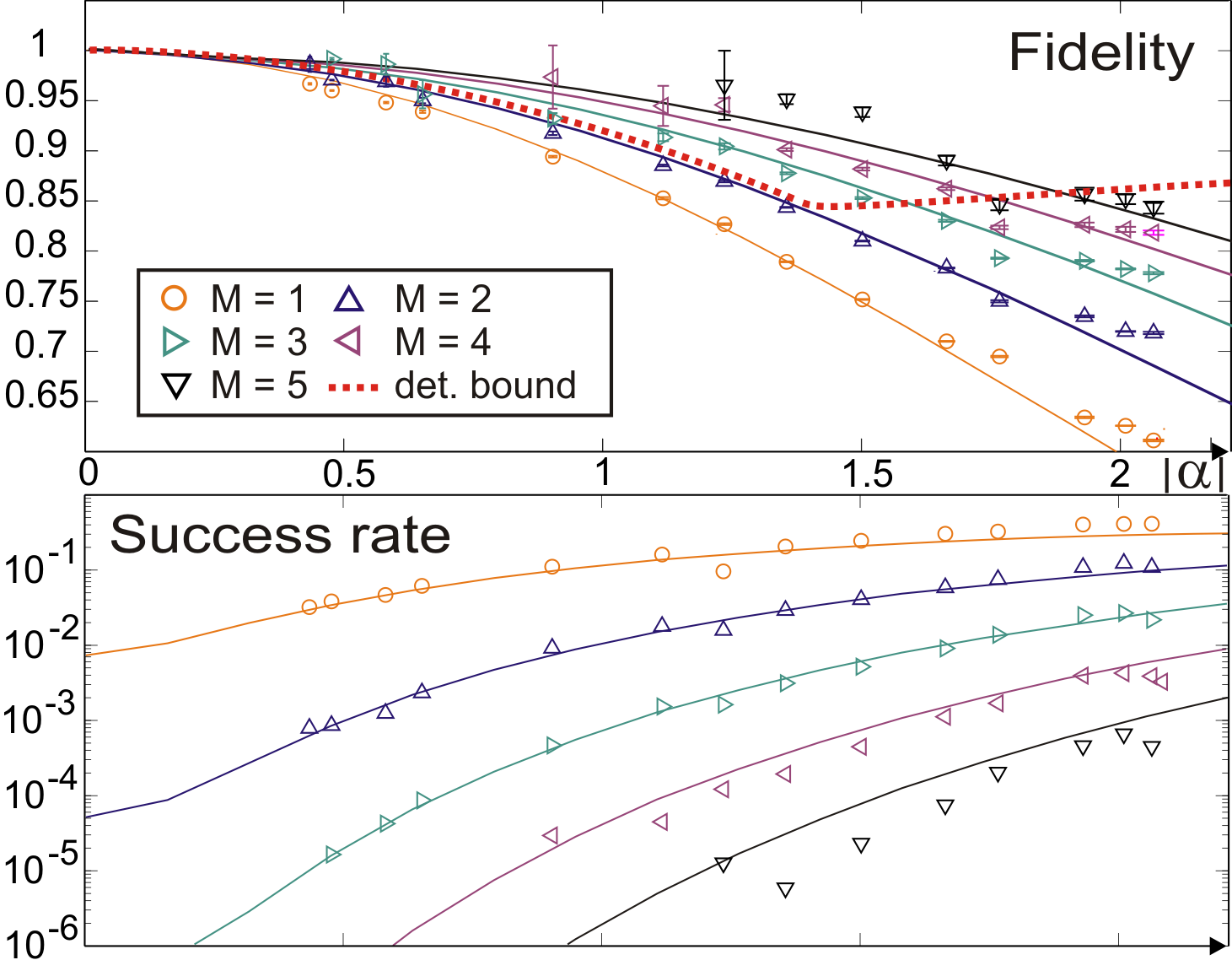}
\caption{(color online) Experimental cloning fidelities and success rates for various input amplitudes and different threshold parameters. The fidelity is maximized in each point over a suitable set of displacement parameters $x$.}
\label{fid_results}
\end{figure}

The primary figure of merit for a cloning device is the fidelity. We measure the average fidelity $F = \frac{1}{2}\langle\alpha|(\rho_{1}+\rho_{2})|\alpha\rangle$ of both clones, to avoid a bias stemming from a possible imbalance of the two outputs. The results for amplitudes in the range of $|\alpha| \in [0.4, 2.1]$ and threshold parameters of up to five photons are shown in Fig.~\ref{fid_results} and are compared to our theoretical model for the implementation with realistic parameters. Additionally, the fidelities achievable with the deterministic scheme from~\cite{Sacchi2007} serve as a threshold. 
For amplitudes of up to $|\alpha|\approx 1.4$ the performance of our probabilistic cloner is comparable to the deterministic scheme at a heralding threshold of $M=2$ and is superior for $M=3$ and above. At higher amplitudes, the fidelities of the deterministic scheme can be surpassed using higher heralding thresholds.
The error bars represent the statistical fluctuations over repeated realizations of the experiment. The measurements were conducted over a period of several weeks, in which variations of up to $\pm 2 \%$ from the average tapping ratio ($17 \%$) occurred due to drifts in the setup.

\begin{figure}
\includegraphics[width=8.5cm]{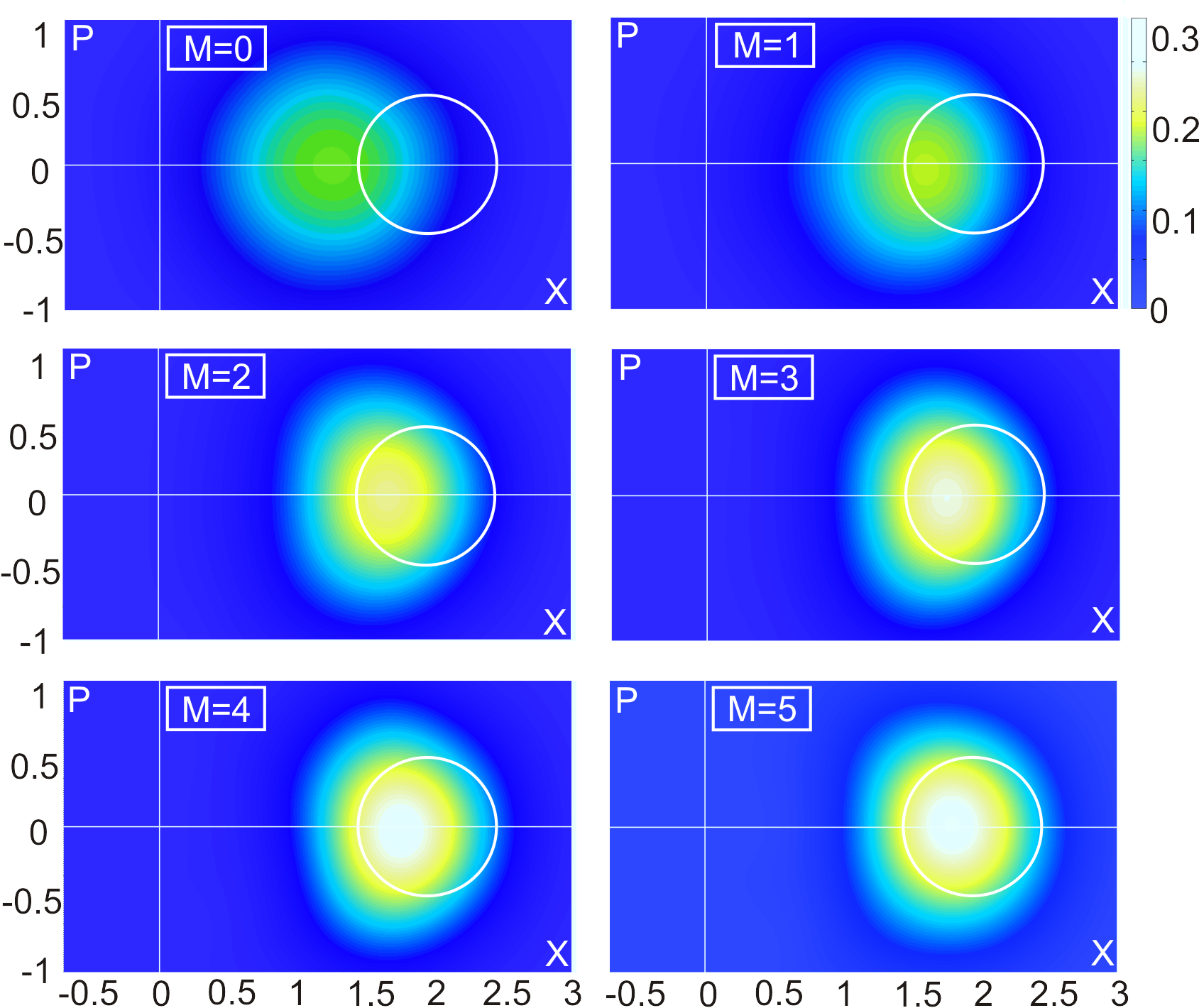}
\caption{(color online) Contour plots of the reconstructed Wigner functions of a single clone for the non-heralded displaced state and for heralding thresholds of up to five detected photons. The coherent input state with $|\alpha|=1.93$ is indicated by the white contour line, corresponding to the height at the standard deviation. The displacement parameter for all $M$ is $x \approx 0.52$. }
\label{WignerContours}
\end{figure}
The success probabilities corresponding to the measurements of the presented fidelities are also shown in Fig.~\ref{fid_results} and compared to the theoretical predictions. A higher threshold parameter and hence an increased fidelity comes at the price of a lower rate of success. However, with rising amplitude the probability to detect a certain number of photons also increases. An example for the experimentally generated clones according to different heralding thresholds is presented as reconstructed Wigner functions in Fig.\ref{WignerContours} for an input state with $|\alpha|=1.93$. In this representation the heralding-induced transition from the randomly displaced state ($M=0$) to a heralded high fidelity clone ($M=5$) can clearly be seen. For just a single clone, our results can be seen as an experimental realization of a quantum preamplifier addressing the problem of channel loss for coherent state communication, which is a problem similar to~\cite{T.C.Ralph_arXiv}.

In conclusion, we have proposed and experimentally realized a cloner based on a probabilistic amplifier with minimal resources. In doing so, we have shown that quantum cloning without phase resources is feasible. 
In good agreement with the theory, we were able to generate high-fidelity clones, beating the hitherto best-performing deterministic approach. We discussed that our scheme allows for a delayed choice between the fidelity and the success rate. Furthermore, the clones exhibit reduced correlations, pointing towards the noiseless nature of the amplification step. 

This work was supported by the Lundbeck foundation and the DFG project LE 408/19-2. R.F. and P.M. acknowledge support from projects ME10156, No. MSM 6198959213 and No. LC06007 of the Czech Ministry of Education, the Grant 202/08/0224 of GA CR and the Alexander von Humboldt Foundation. P.M. acknowledges support from the Grant P205/10/P319 of GA CR.

\end{document}